# Supervised Multi-Regional Segmentation Machine Learning Architecture for Digital Twin Applications in Coastal Regions

Mohsen Ahmadi[1], Ahmad Gholizadeh Lonbar[2,3], Mohammadsadegh Nouri[4,3], Amir Sharifzadeh Javidi[3], Ali Tarlani Beris[5,6], Abbas Sharifi[7,*], Ali Salimi-Tarazouj[8,9,10]

1- Department of Electrical Engineering and Computer Science, Florida Atlantic University, FL, USA
2- Department of Civil, Architectural and Environmental Engineering, Missouri University of Science and Technology, Rolla, MO, USA
3- School of Civil Engineering, College of Engineering, University of Tehran, Iran
4- Center for Applied Coastal Research, Department of Civil and Environmental Engineering, University of Delaware, Newark, DE, USA
5- Department of Mechanical Engineering, College of Engineering, Boston University, Boston, MA, USA
6- School of Mechanical Engineering, College of Engineering, University of Tehran, Iran
7- Department of Civil and Environmental Engineering, Florida International University, Miami, FL, USA
*Corresponding author: asharifi@fiu.edu*
8- NWS/NCEP/Environmental Modeling Center, National Oceanic and Atmospheric Administration (NOAA), College Park MD, USA
9- Lynker, Leesburg, VA 20176, USA
10- University of Maryland, College Park, MD 20742, USA

**Abstract**

This study explores the use of a digital twin model and deep learning method to build a global terrain and altitude map based on USGS information. The goal is to artistically represent various landforms while incorporating precise elevation modifications in the terrain map and encoding land height in the altitude map. A random selection of 5000 segments from the worldwide map guarantees the inclusion of significant characteristics in the subsets, with rescaling according to latitude accounting for distortions caused by map projection. The process of generating segmentation maps involves using unsupervised clustering and classification methods, segmenting the terrain into seven groups: Water, Grassland, Forest, Hills, Desert, Mountain, and Tundra. Each group is assigned a unique color, and median filtering is used to improve map characteristics. Random parameters are added to provide diversity and avoid duplication in overlapping image sets. The U-Net network is deployed for the segmentation task, with training conducted on the seven terrain classes. Cross-validation is carried out every 10 epochs to gauge the model's performance. The segmentation maps produced accurately categorize the terrain, as evidenced by the ROC curve and AUC values. The main goal of this research is to create a digital twin model of Florida's coastal area. This is achieved through the application of deep learning methods and satellite imagery from Google Earth, resulting in a detailed depiction of the coast of Florida. The digital twin acts as both a physical and a simulation model of the area, emphasizing its capability to capture and replicate real-world locations. The model effectively creates a global terrain and altitude map with precise segmentation and capture of important land features. The results confirm the effectiveness of the digital twin, especially in depicting Florida's coastline.

# 1. Introduction

The development of deep learning in recent years has resulted in its increasing use across a wide range of application areas. Deep learning is currently gaining rapid traction in remote sensing due to its success and the increasing availability of data and processing resources. In addition to playing a significant role in the government's scientific control of aquaculture resources, coastal aquaculture areas are also prone to storm tide disasters, which is a characteristic of remote sensing sites. There is an increasing number of researchers who are exploring ways to extract information from aquaculture areas using remote sensing and machine learning technologies, which has led to a number of follow-up studies [1-4]. In order to determine aquaculture zones, researchers use threshold segmentation [5, 6], semantic segmentation networks [7], feature learning [8-10], expert experience [11,12], and feature learning. A variety of artificial intelligence techniques are used in the Digital Twin Framework in order to analyze remote sensing-derived images and point cloud models. As part of the process of collecting data on land use and land cover, image segmentation is the first step, in which remote sensing images are divided into categories based on their semantic content, such as urban areas, agricultural fields, and woods. As part of the damage assessment process, object detection and tracking algorithms can also be used to identify major infrastructure items such as bridges and highways [13]. A method of change detection is used to identify changes in the built or natural environment caused by environmental disasters such as floods or earthquakes by comparing two geospatially coordinated remote sensing images. In disaster response and emergency management, artificial intelligence may be able to enhance crowdsourcing's efficiency by providing high-quality, structured data that can be used to train machine learning models. The use of crowdsourcing involves combining the contributions of many individuals to improve the quality of decisions [14]. A crowdsourcing approach to disaster management primarily involves the collection and processing of data [15]. By using it, individuals can perform tasks that are easy for humans but complicated for machines, such as classification or labeling images, and communicating information regarding disasters in real-time. Therefore, structured data of high quality is produced, which facilitates machine learning algorithm training and decision-making. Remote sensing images and social media posts can be used to identify damaged coastal areas. The use of crowdsourcing facilitates the analysis of remote sensing and social sensing data by providing tagged images for machine learning model training within the Digital Twin framework, as well as providing high-quality information regarding crisis scenarios [16].

Crowdsourcing technologies could improve the efficiency of machine learning algorithms like Human-in-the-Loop (HITL) [17]. The development of digital twins has expanded our knowledge of real-world events and improved our modeling capabilities. Digital twins enable us to model and analyze complex systems, thus facilitating informed planning and decision-making [18]. A digital twin is fashioned by computer simulations that mimic real-world phenomena. In this study, we illustrate the application of deep learning techniques to create a digital twin for the Florida coastline. This coastal region, characterized by its dynamism and biological significance, features a diverse terrain including water bodies, grasslands, forests, hills, deserts, mountains, and tundra. To accurately depict this region's physical traits from satellite data, we used deep learning methods, including the U-Net network, for segmentation. By employing satellite imagery and the U-Net network, we compiled comprehensive data on the varied terrain of the Florida coastline. This segmentation data forms the cornerstone of the digital twin, enabling us to accurately portray the geographic distribution and features of each terrain type. The deep learning-based segmentation process can generate binary masks for each terrain type, delineating boundaries between different land covers and between land and water. Incorporating these masks into the digital twin model will simulate the coastal region's behavior over time, assisting us in comprehending its dynamics. Deep learning-based segmentation can also monitor changes in land cover, such as those triggered by urbanization, deforestation, and coastal erosion. By comparing satellite images from various time periods, we can detect and quantify these changes, offering valuable insight for land management, environmental conservation, and urban planning. Future enhancements and improvements to the digital twin model may involve applying deep learning techniques and integrating additional data sources like climatic or hydrological data. This allows for modeling environmental impacts on the coastal region, furthering our understanding of its dynamics and aiding in projecting probable future scenarios.

Digital twins have applications beyond mere modeling and simulation. A digital twin of the Florida coast can impact multiple industries including tourism, environmental management, urban development, and disaster preparedness. Urban planners can leverage the digital twin to visualize and evaluate potential future growth scenarios, considering environmental implications and promoting sustainable development. Environmentalists can employ the digital twin for monitoring ecosystem health and advancing conservation efforts. Emergency management organizations can use the digital twin to mitigate the impacts of natural disasters through preparedness and response planning. The innovative design of the digital twin can also enable visitors and local communities to conduct virtual explorations and develop sustainable tourism strategies that benefit all parties. With deep learning methods, notably the U-Net network, we were able to accurately segment and represent terrain types in creating the digital twin of the Florida coast. This approach offers a deeper understanding of the physical characteristics and dynamics of coastal regions, facilitating informed decision-making and planning. Digital twins offer valuable insights and opportunities in various fields, thereby contributing to sustainable development and resource management.

## 2. Related work

Despite the impressive performance of pre-trained Convolutional Neural Networks (CNNs), their need for extensive training data and potentially insufficient resilience limits their applicability across different contexts [19,20] (see Table 1). In contrast with image classification and object recognition tasks, semantic segmentation requires precise per-pixel annotations for each training set. Semantic segmentation networks [21,22] require a significant volume of labeled data to interpret images at the pixel level. Several authors have adopted the Multiple Instance Learning (MIL) method for creating labels in supervised learning, including Pinheiro and Collobert [23], and Pathak et al. [24]. Souly et al. [25] suggested a semi-supervised learning technique, later refined by Hong [26], that utilized unlabeled or weakly labeled data as well as artificially created GAN images. Hu et al. [27] developed a model using a modified YOLO-V4 network for identifying uneaten feed particles in aquaculture, successfully addressing issues such as low contrast, small size, and abundance of underwater feed particles in images. According to Moazzam et al [28], a deep learning-based semantic segmentation model was employed to identify weeds in agriculture. By training an improved U-Net-based backbone feature extraction network, they managed to visually discern crop weeds in the field by distinguishing between the foreground and background. However, with deep learning-based image segmentation algorithms, it's necessary to precisely label many samples to enhance the model's prediction accuracy. Ensuring labeling accuracy in pellet images can be challenging, as the edge and background distinction is not clear. Bousmalis et al. [29] employed a GAN-based approach to adapt source domain images to the target domain, yielding convincing results and outperforming several unsupervised domain adaptation scenarios. Xue et al. [30] used a fully convolutional neural network as a segmenter in their SegAN network, training the critic and segmenter with a multi-scale L1 loss function.

**Table 1.** Components of Digital Twin Simulation for Disaster City and Artificial Intelligence Methods

| Components | Stages of a Disaster | Method | Deficiencies/possibilities | Processes that Artificial Intelligence can enhance |
|---|---|---|---|---|
| **Employing multi-sensor data gathering methods** | Readiness; action; and restoration | Aerial images captured by satellites and drones; spreading of information; and machine learning with human involvement | Image segmentation, misinformation identification, and data annotation | Swiftly gather details about the situation, gauge the magnitude of calamities, and chart the destruction in impacted regions |
| **Data consolidation and examination** | Reaction; restoration; and prevention | Conceptual networks; network integration; and categorization of entities | Utilizing a projection function for decreasing dimensionality and eliminating redundancy | Examine the crisis scenario in-depth and determine the connections among the relevant entities involved |
| **Decision making based on game theory involving multiple participants** | Reaction; and restoration | AlphaZero strategy; Monte-Carlo decision tree exploration | Simulation of decision-making procedures; and task diagram creation | Examine the interplay between various elements and the uncertainties stemming from the involvement of numerous actors |
| **Analysis of evolving network structures** | Readiness; action; and restoration | Meta-networks; and link prediction. | Determine connections between participants; and forecast changes within network structures | Determine essential actors and evaluate the effectiveness of emergency response procedures |

Liang et al. [31] introduced the Sem-GAN framework that notably enhanced the quality of translated images. Building upon the methodology of CGAN and previous research, they developed Semi-SSN, a novel network that integrates conditional adversarial learning into semantic segmentation networks, supporting semi-and weakly-supervised learning. The model's training involves the generation of pseudo-labels through a combination of confidence maps created by the discriminator and forecasted maps by the generator of unlabeled data. To mitigate the significant disparity between source and target data, unsupervised domain adaptation frameworks have been established. Zou et al. [32] introduced a self-training technique with class balance. Li et al. [33] devised a bidirectional model and a self-supervised learning system for concurrent domain adaptation and image segmentation. Luo et al. [34] suggested a category-level adversarial network that merges co-training and adversarial learning concepts, aiming to maintain local semantic consistency during global feature alignment. Using a geometry-consistent GcGAN and a co-training adversarial network, Fang et al. [35] developed a category-sensitive technique for mapping land cover using optical aerial images. Traditional transfer learning models can effectively utilize remote sensing images, even though many were specifically designed for natural images. These techniques extract high-level semantic information from remote sensing images using unique procedures. However, GcGAN-based models often face instability in convergence processes and inadequate training due to their complex structural designs. Castillo-Martinez et al. [36] developed a color index-based thresholding method for segmenting plant images under controlled conditions, using a fixed and automated thresholding approach. However, a threshold segmentation method might struggle to find the ideal threshold value. A too high threshold might lead to the loss of target areas, while a too low threshold may increase background areas. This method is also sensitive to noise, making it unsuitable for segmenting images with significant noise. In coastal remote sensing, innovative techniques for semantic segmentation have been suggested in the literature study. There are several problems with these techniques, including the lack of labeled data, complicated land cover details, and weak model robustness. For applications such as land-use mapping, aquaculture extraction, and urban waterlogging, precise and effective segmentation is crucial [37].

According to Chen et al. [38] a semi-supervised semantic segmentation system based on pseudo supervision was proposed with emphasis on feature representations and perturbation applied to impose consistency restrictions. In contrast to conventional consistency training, they employed predictions from an additional segmentation network for online pseudo supervision. When mapped land use and land cover in semi-supervised circumstances, the framework demonstrated satisfactory performance. The issue of finding high-quality pre-labeled training samples for coastal remote sensing images segmentation was addressed by Fang et al. [39]. They proposed the conditional co-training (CCT) technique for genuinely unsupervised segmentation. The CCT framework consisted of two simultaneous data streams: pixel-level semantic segmentation and super pixel-based over segmentation. By employing various conditional constraints to extract high-level semantic information, the suggested method generated full-resolution segmentation maps without pre-labeled ground truths. In comparison to other unsupervised approaches, CCT demonstrated plausible performance and superior efficacy. For the scientific management and planning of aquaculture resources, Liang et al. [40] have focused on extracting coastal aquaculture sites. Based on three different sources of data, they proposed a new semi-/weakly-supervised semantic segmentation network (Semi-SSN) using a variety of spatial resolutions. Among the issues addressed by the Semi-SSN approach were problematic labeling, weak model resilience, and variable spatial resolutions. Using extensive trials and comparisons with cutting-edge techniques, the authors demonstrated the efficacy of Semi-SSN in extracting coastal aquaculture regions. According to Tzepkenlis et al. [41], an effective semantic segmentation strategy for satellite image time series can be modified by combining single multiband image composites with temporal median filtering. In order to improve land cover classification performance while reducing complexity, denoised composites were used. The authors evaluated several combinations of satellite data and recommended using channel attention rather than temporal attention. Despite using fewer training parameters, the improved strategy performed significantly better than other well-liked methods in terms of mean intersection over union (mIoU). Using an innovative model combining extreme gradient boosting (XGBoost) and semantic segmentation of satellite images, Tong et al. [42] used extreme gradient boosting (XGBoost) to assess urban waterlogging risk in coastal cities in 2023. In satellite images, the semantic segmentation model successfully located water bodies, roads, and greenery, whereas XGBoost successfully predicted waterlogging spots and identified the main causes of urban waterlogging. According to the full performance evaluation of the models, elevation has the greatest impact on waterlogging, in addition to secondary elements such as roads and water bodies. According to Zhou et al. [43], unbalanced data can make it difficult to segment remote sensing images. They provide a dynamic weighting technique and an efficient sample computation technique to help increase segmentation accuracy. Based on experimental data, minimal-class segmentation has demonstrated significant improvements in accuracy and recall for identifying a forest fire's burning area.

## 3. Methods and Materials

### 3.1. *Convolutional Neural Network (CNN)*

Deep learning, a subset of machine learning, employs several non-linear transformations to attain higher-level abstraction. It's widely used in computer vision and remote sensing, with one common use being image classification through the application of convolutional neural networks (CNNs). CNNs employ a form of multiplication known as convolution, making them a prevalent deep learning technique [44]. Lately, CNNs have seen extensive use in classifying land cover images. A standard CNN comprises of convolutional layers, pooling layers, and fully connected layers, which form the fundamental building blocks of its structure. A typical arrangement might involve a series of stacked convolutional layers, succeeded by a pooling layer, and then a fully connected layer [45, 46]. The process of forward propagation transforms the input data into the output through these layers. Although this section discusses the convolution and pooling methods for a two-dimensional (2D) CNN, the same operations can also be performed with a three-dimensional (3D) CNN.

Feature extraction, a blend of linear and non-linear processes, specifically convolution and activation, is carried out by the crucial convolution layer in CNNs. Max pooling is the most commonly used pooling operation. Typically, a max pooling technique uses a filter size of 2x2 and a stride of 2, resulting in a two-fold down sampling of the feature dimensions. The algorithm selects patches from the input feature maps, keeps the highest value from each patch, and throws away all other values. The depth dimension of the feature maps does not vary, in contrast to height and breadth. Various regularization techniques have been proposed for deep neural networks, including batch normalization, weight decay, data augmentation, among others, with the dropout regularization technique being the most used. The dropout method serves as an adaptive regularization technique, equivalent to the L2 regularizer in first order for generalized linear models. Furthermore, dropout adaptively adjusts the regularization strength in response to the inputs and variances of the dropout variables. The transposed convolutional layer is sometimes referred to as a deconvolutional layer, contrasting with a standard convolutional layer. Unlike a standard convolutional layer, a regular convolutional layer reproduces the original input when deconvolved. The spatial dimensions are the same in both transposed convolutional and deconvolutional layers. Transposed convolution inverts the dimensions of the convolution, unlike conventional convolution which works on values.

### *3.2. U-Net Architecture*

The U-Net structure, developed for image segmentation utilizing CNNs, was introduced by Ronneberger et al. [47] and has since been applied extensively in various fields. In the U-Net architecture, encoders and decoders play an important role. Convolutional and pooling layers are applied to the input image to reduce the spatial resolution and increase the number of feature maps. Assimilation of the overall context of the image is carried out by the encoder, which extracts high-level features. The decoder, on the other hand, gradually increases the size of the encoder's output, decreasing the number of feature maps while restoring the size of the encoder's output to its original size.

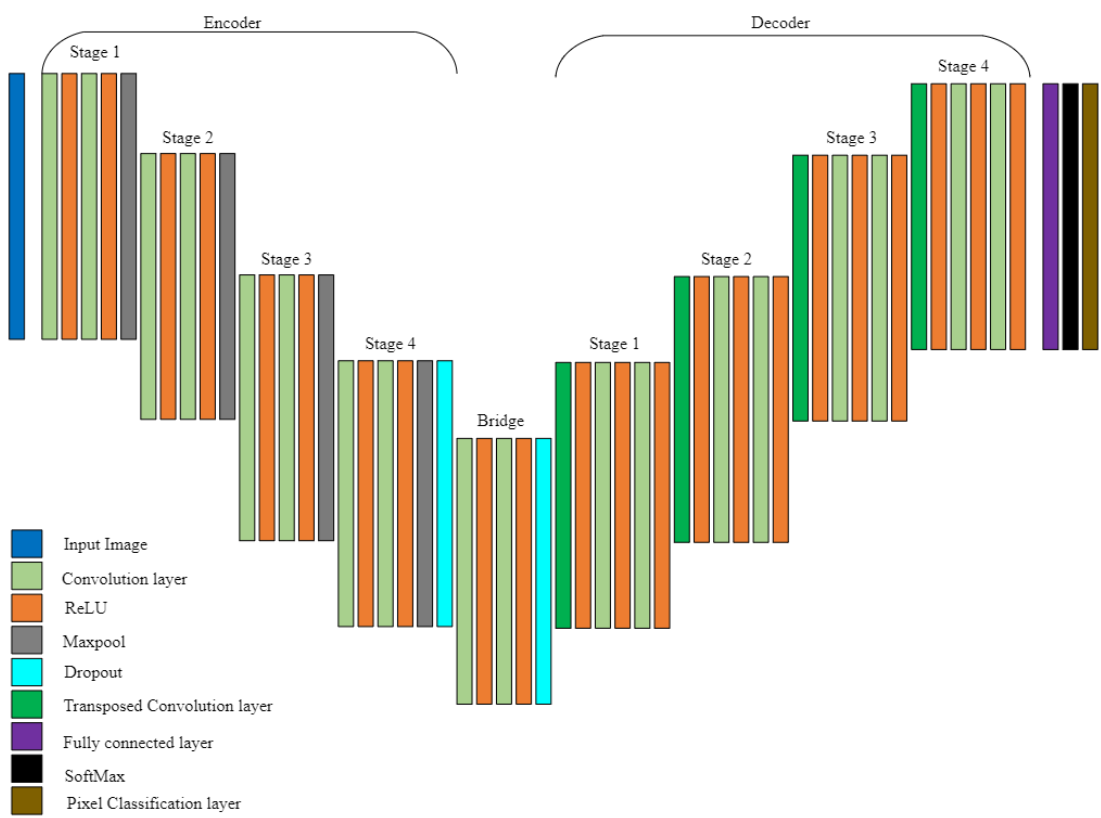

**Fig. 1**  Network architecture

Segmentation maps are created by combining high-level features from the encoder with precise spatial information from low-level features. The convolutions are typically adjusted or interpolated during this magnification process. There is also a notable feature of U-Net architecture in the presence of skip connections between encoders and decoders at matching levels. This allows the decoder to access low-level features from the encoder directly, enhancing segmentation's accuracy and preserving spatial data. Figure 1 depicts the design of the U-Net architecture.

# 4. Results and Discussion

## *4.1. Data Collection*

The data used in this study was collected from the United States Geological Survey (USGS) dataset. The dataset consisted of a global Terrain and Height map, which provided information about the land type, elevation changes, and altitude above sea level. The Terrain map was an artistic representation of land features, incorporating geometrically calculated shadowing and highlights to depict changes in elevation. The Height map encoded land altitude as pixel intensity. To create a representative dataset for analysis, a total of 5000 patches were randomly selected from the global map. Care was taken to ensure that the selected patches contained meaningful features rather than being solely comprised of empty ocean regions. Additionally, the crops were rescaled based on latitude to account for the distorting effects of the map projection and to maintain consistent sizes for terrain features across different latitudes. The generation of the dataset involved employing unsupervised clustering and classification techniques for local pixel regions. This process allowed for the segmentation of the patches into distinct terrain categories. In total, seven terrain categories were defined for segmentation, each associated with specific representative colorings. The terrain categories and their corresponding representative colorings were as follows:

1. Water: RGB color code (17, 141, 215)
2. Grassland: RGB color code (225, 227, 155)
3. Forest: RGB color code (127, 173, 123)
4. Hills: RGB color code (185, 122, 87)
5. Desert: RGB color code (230, 200, 181)
6. Mountain: RGB color code (150, 150, 150)
7. Tundra: RGB color code (193, 190, 175)

Furthermore, to enhance the visual quality of the segmentation maps and reduce noise, a median filtering technique was applied. This process smoothed out the features, resulting in larger, more coherent regions or blobs within the maps. Each segmentation map was created with randomized parameters to introduce variety across the dataset and ensure that any overlapping image sets would not have identical segmentation maps. By following these data collection procedures, a diverse and representative dataset was assembled, enabling comprehensive analysis of the global Terrain and Height maps.

## *4.2. Results of training process of Segmentation Method*

The results of our thorough investigation into the building of a worldwide Terrain and Height map and the production of a digital twin for the Florida coastal region using cutting-edge deep learning methods are presented in this section. Our goal was to utilize deep learning techniques to depict and categorize different types of landscapes. Initially, we created a worldwide Terrain and Height map by leveraging information from the USGS. The primary objective of the Terrain map was to provide a visually appealing representation of diverse terrain types, while also accurately capturing variations in elevation through the effective use of shadows and highlights. To ensure the inclusion of significant features, we randomly selected 5000 patches from the world map to represent the height of land regions above sea level using pixel intensities. As part of our efforts to mitigate any distortions resulting from map projections, we also rescaled the patches based on their respective latitudes, thereby maintaining consistent topographical characteristics.

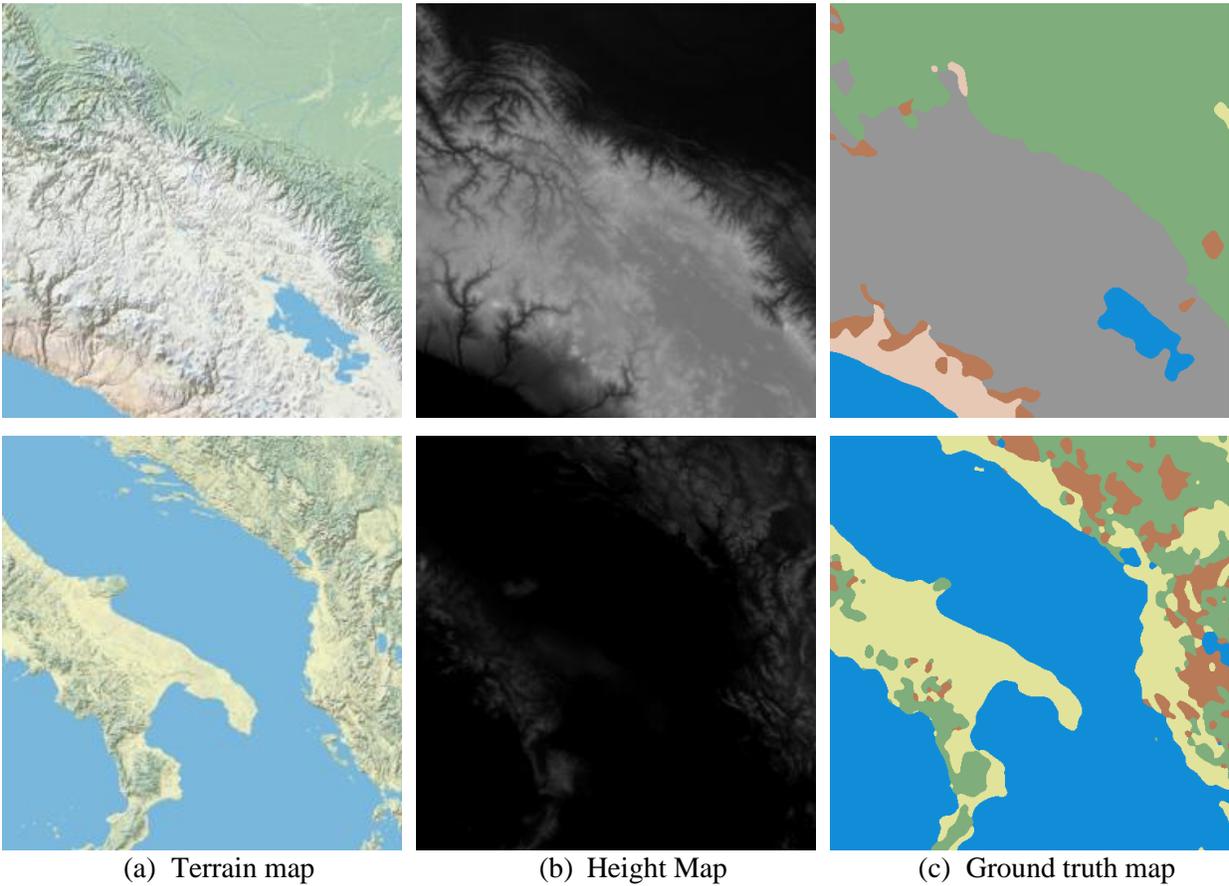

(a) Terrain map       (b) Height Map       (c) Ground truth map

**Fig. 2** Representationg of dataset, a) original terrain map, b) hight map, c) grountruth of the area

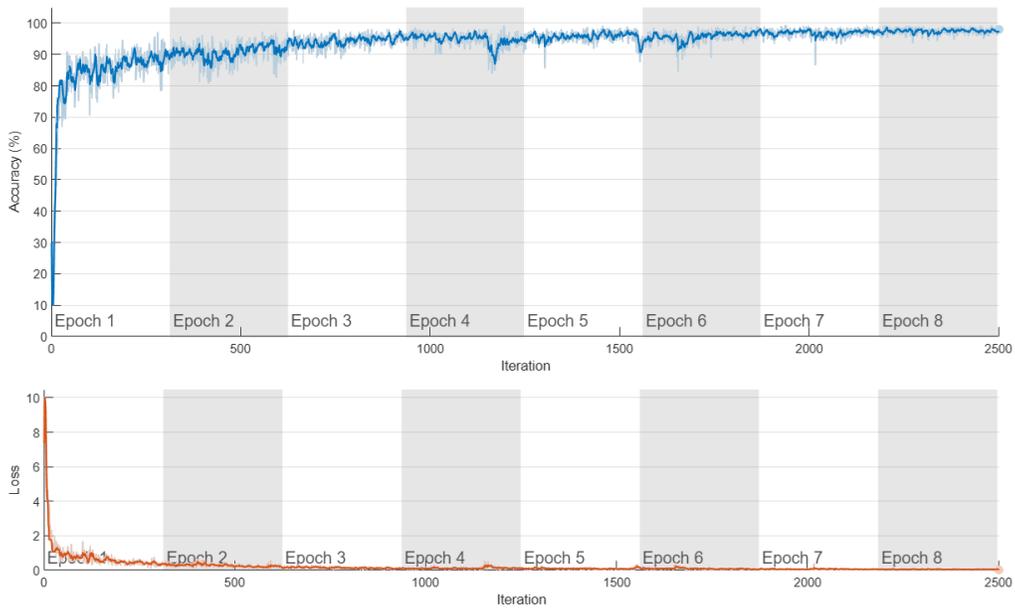

**Fig. 3** Training process of the provide deep leanirng network

In addition to Water and Grassland, the terrain categories encompassed Forest, Hills, Desert, Mountain, and Tundra. To enhance the quality of the segmentation maps and reduce noise, a median filtering technique was applied, resulting in smoother and more cohesive features within the maps.

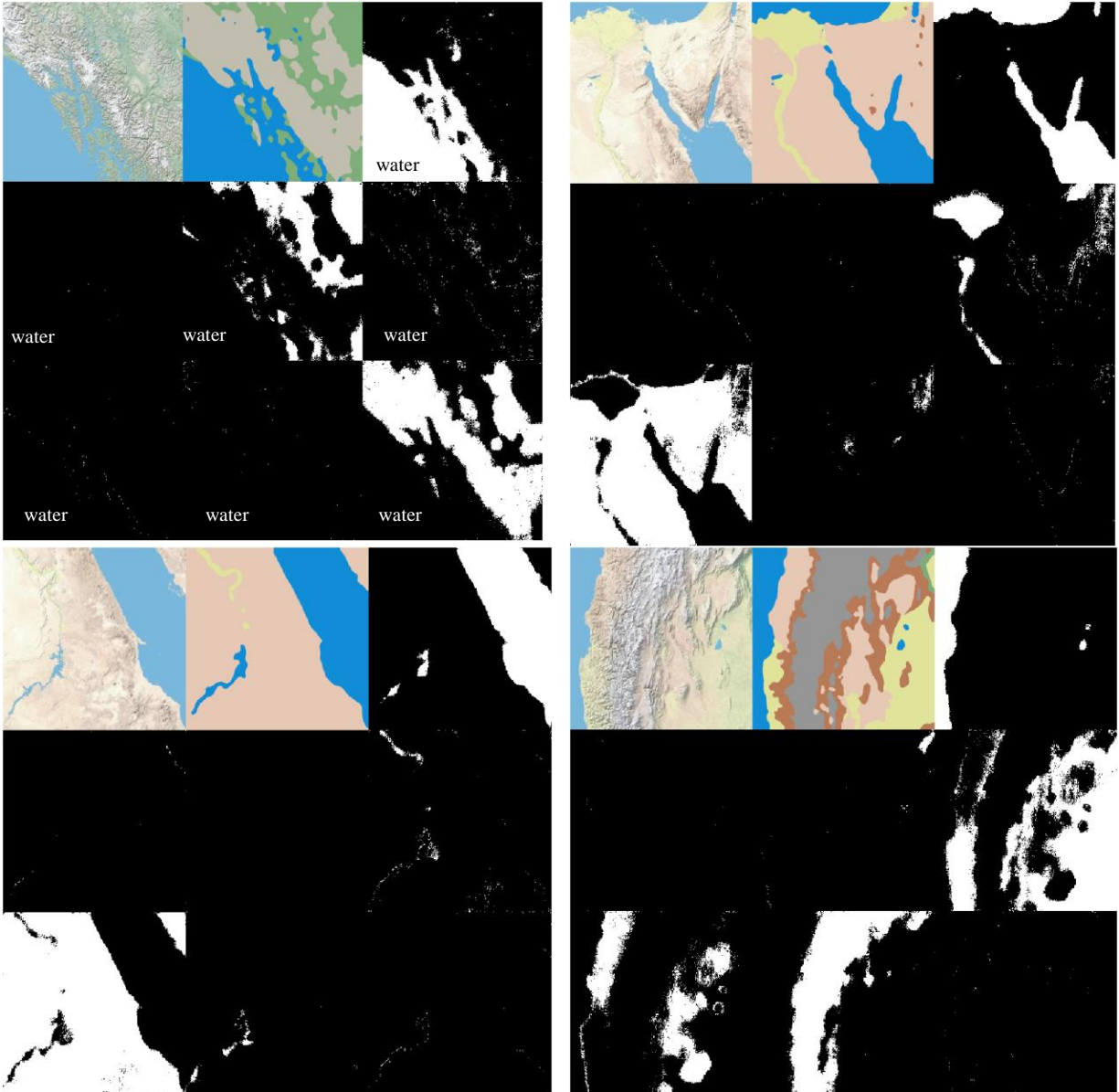

**Fig. 4**  Results of segmentation

This process generated large and coherent blobs that represented the segmented regions effectively. To ensure diversity and avoid redundancy in the dataset, the creation of segmentation maps incorporated randomized parameters. Consequently, overlapping image sets did not possess identical segmentation maps. For the segmentation task, the U-Net network, a deep learning architecture, was employed. The network was trained on the seven terrain classes using a maximum of 300 epochs, a minibatch size of 16, and a learning rate of 1e-4. The model's performance was evaluated through cross-validation conducted every ten epochs. Figure 3 provides a visual representation of the model's progress based on accuracy and loss values throughout the training process. Figures 4 and 5 present the outcomes of the segmentation process utilizing the deep learning-based methodology.

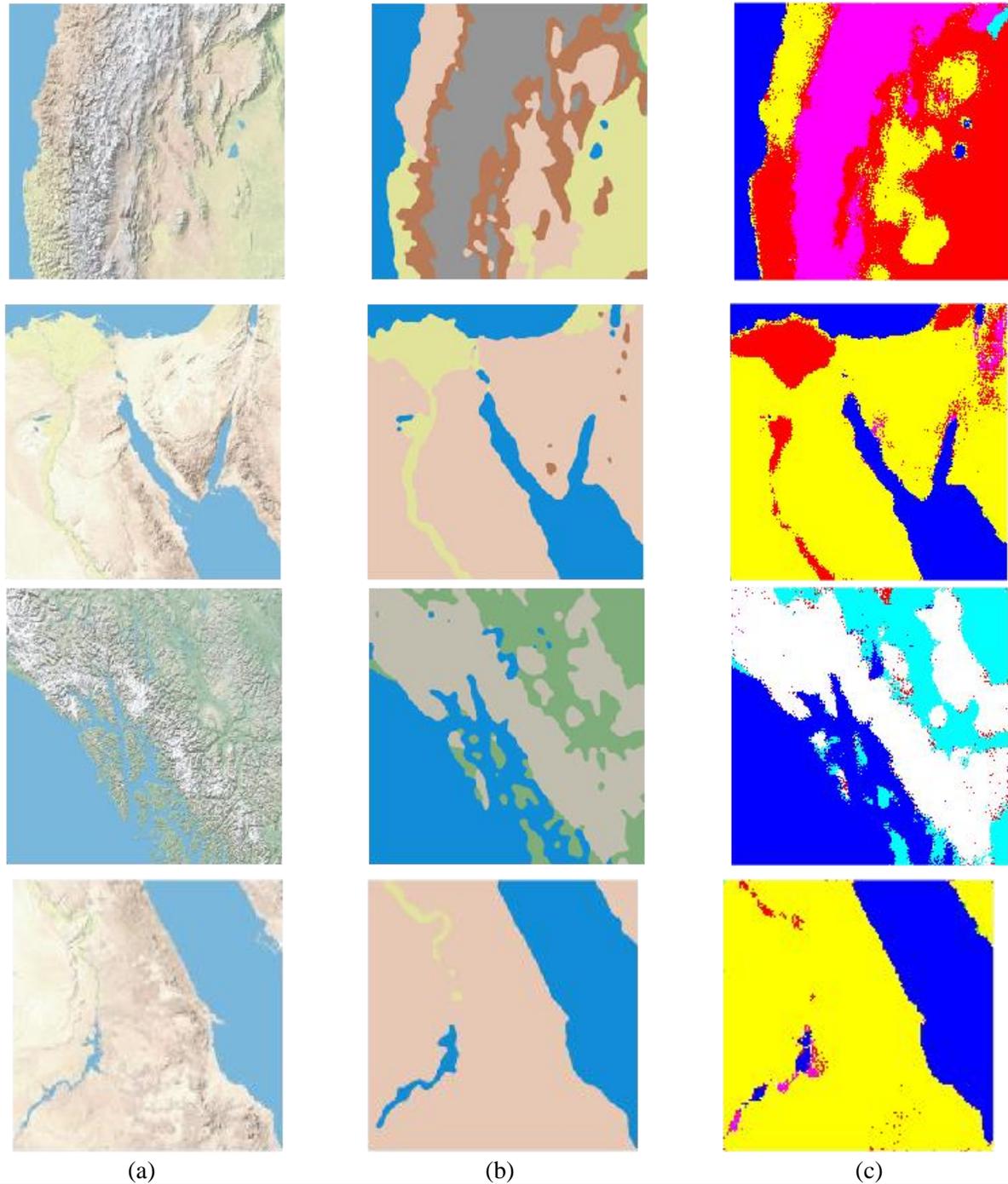

| (a) | (b) | (c) |

| Background (Black) | Water (Blue) | Grassland (Green) | Forest (Cyan) | Hills (Red) | Desert (Yellow) | Mountain (Magenta) | Tundra (White) |

**Fig. 5** Results of presented deep learning method for detection of environmental area

These figures serve as visual representations of the segmentation process, showcasing binary masks corresponding to different terrain types. Figure 1 provides a comparison between the original image and the ground truth image, which serves as a reference for evaluating the segmentation accuracy. Subsequent images in the figure display binary masks that accurately identify the boundaries of various landscape types, including water, grassland, forests, hills, deserts, mountains, and tundra. The generation of these binary masks plays a crucial role in creating realistic and visually appealing representations of the digital twin. By leveraging these masks, viewers can easily identify and visualize distinct terrain types within coastal regions, gaining a clear understanding of their spatial distribution and unique characteristics

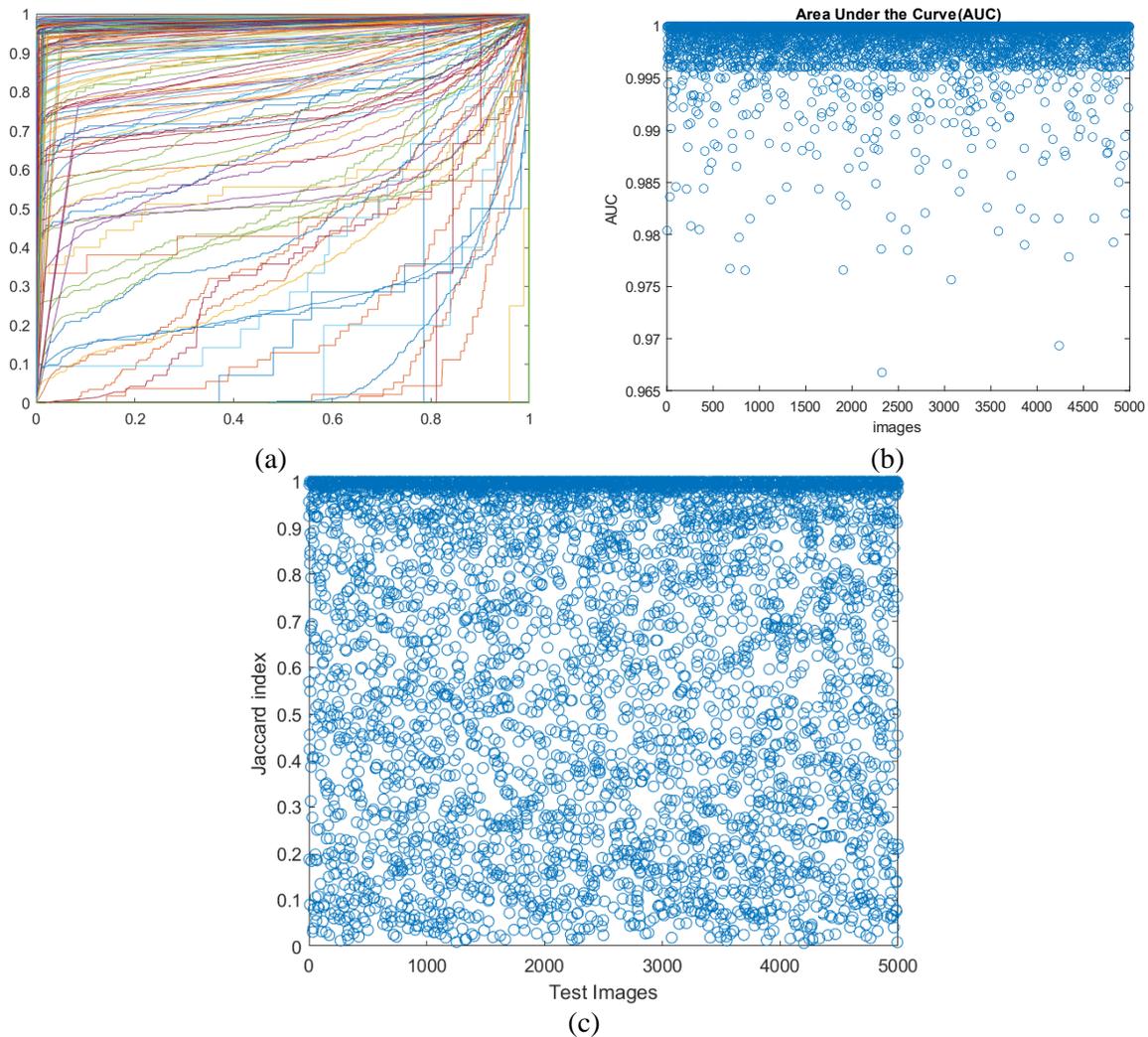

**Fig. 6**   Performance metrics a) ROC curve, b) AUC value c) Jaccard index

To assess the performance of our segmentation model, we conducted a comprehensive evaluation using a receiver operating characteristic curve (ROC), as illustrated in Figure 6a. The ROC curve provides insights into the model's ability to detect various terrain categories based on the images from the test dataset. Additionally, Figure 6b presents the calculated values for the area under the curve (AUC), which serve as a quantitative measure of the model's accuracy in differentiating terrain categories.The analysis revealed that a significant proportion of images achieved high AUC values, indicating accurate and precise segmentation. Moreover, to measure the similarity between the ground truth and the segmentation results obtained from the U-Net network, we employed the Jaccard index. The Jaccard index allows for a quantitative assessment of the overlap or similarity between two sets of data. By utilizing this index, we were able to validate the fidelity of the segmentation results in comparison to the ground truth.

### *4.3. Visual representation Florida state*

Figure 7 presents a satellite image of the entire Florida state on a scale of 200 km. This image provides a broad overview of the entire region, allowing us to visualize the extensive coastline and surrounding land areas. The satellite image serves as the foundational data source for our digital twin modeling efforts.

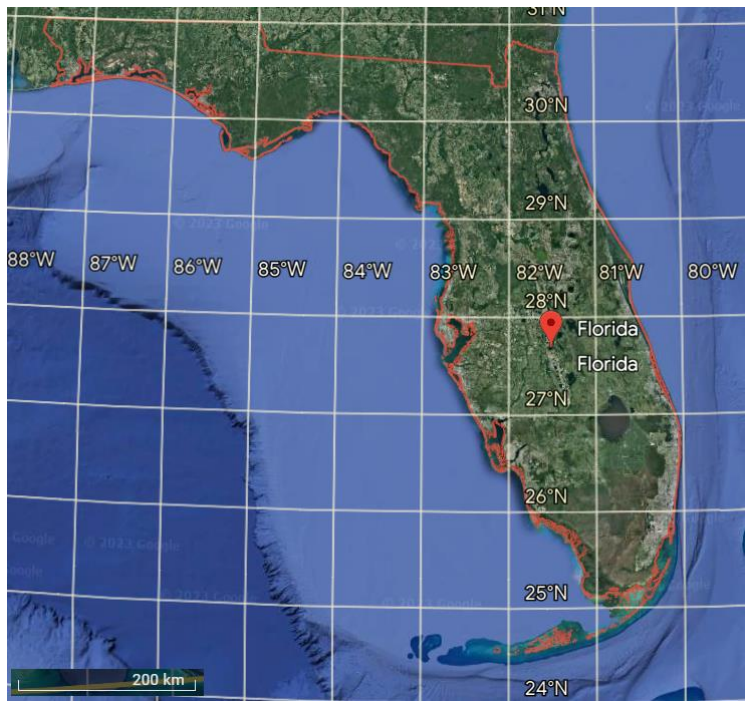

**Fig. 7**  Florida satellite image based on Google Earth

To focus specifically on the coastal region of Florida, we zoomed in and divided a smaller area measuring 10 km in size. This approach allowed us to concentrate on capturing the detailed features of the coastline and its surrounding rural areas, ensuring an accurate portrayal within the digital twin. Subsequently, we applied the segmentation model to this 10 km coastline area that was previously discussed in earlier sections of our research. The segmentation process effectively categorized pixels in the area into various types of topography, including water, grassland, forests, hills, deserts, mountains, and tundra.To create a comprehensive digital twin of the Florida coastline region, we downloaded satellite images from Google Earth at multiple resolutions. Leveraging these images, we were able to faithfully recreate the distinctive characteristics and traits of the Florida coastline region. The outcome of this modeling approach is showcased in Figure 8, which presents an exceptionally detailed digital twin of the Florida coastline region, showcasing the successful implementation of our methodology

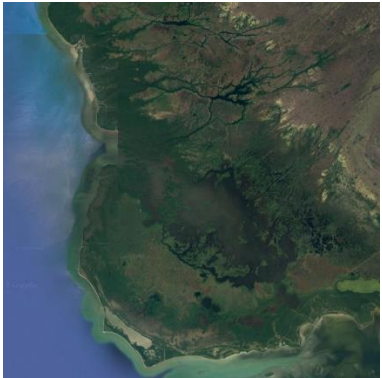
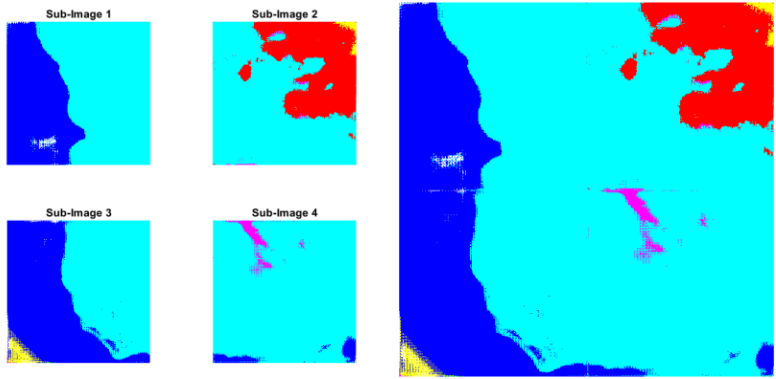

Everglades National Park

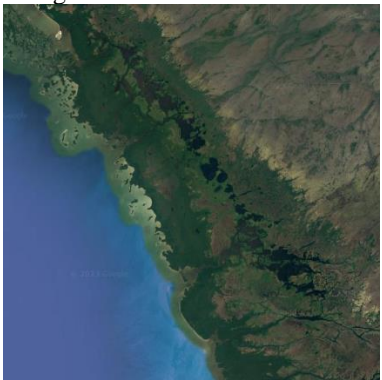
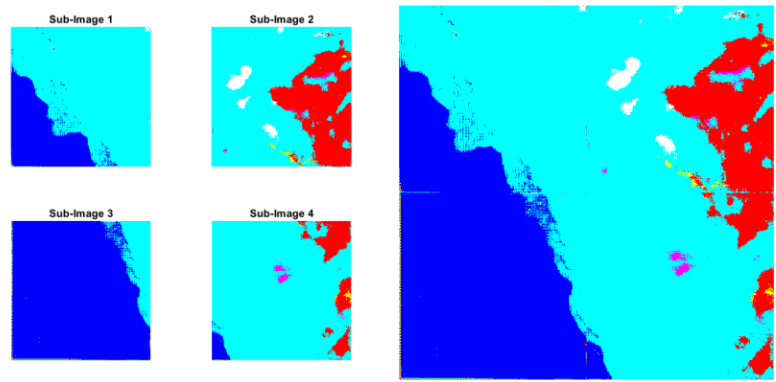

Big Cypress wildlife management area

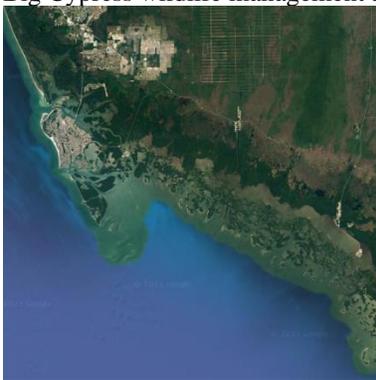
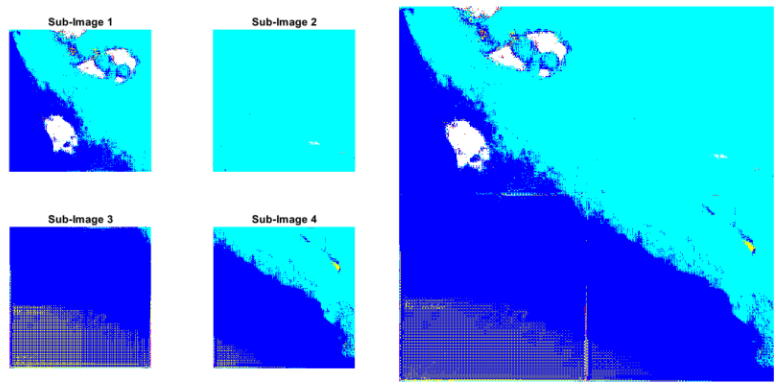

Cape Romano Islands

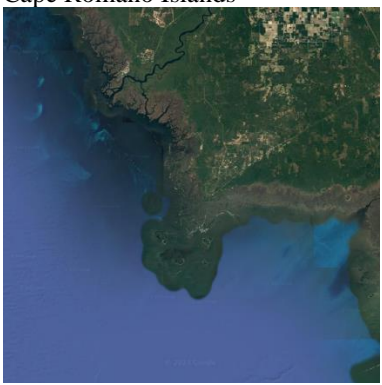
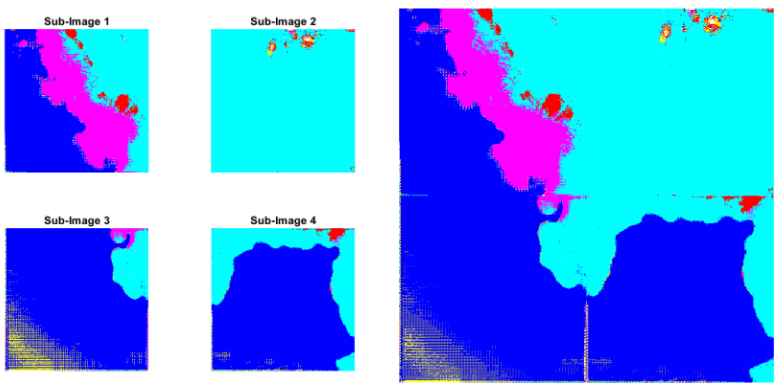

Waccasassa Bay Preserve state park

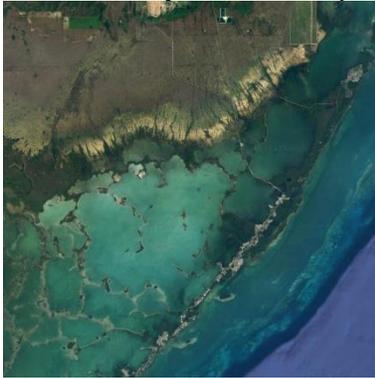
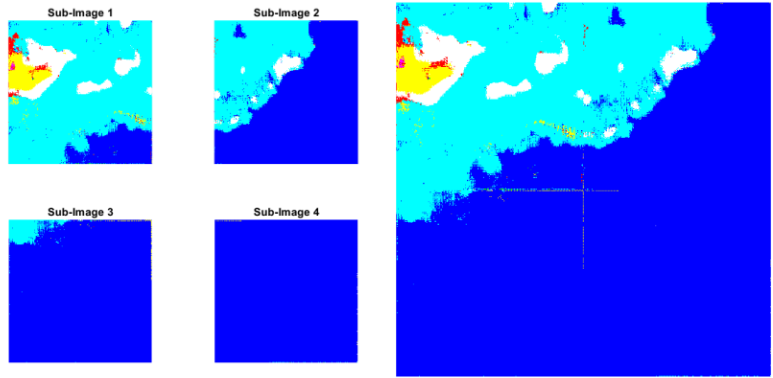

Southern Glades wildlife and Environmental area

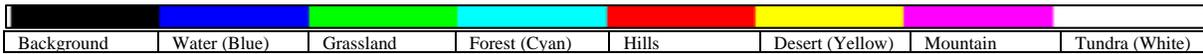

| Background | Water (Blue) | Grassland | Forest (Cyan) | Hills | Desert (Yellow) | Mountain | Tundra (White) |

**Fig. 8** The visual representation of some wild life parks in Florida detected by the presented digital twin model

By merging the segmented data from the 10 km coastal area with the larger satellite image of the entire state of Florida, we achieved a detailed and comprehensive depiction of the coastline region within the digital twin. This integration allowed for a seamless transition between the overall perspective of the state and the finer-scale details of the coastal area. By utilizing satellite images at various resolutions and employing our segmentation approach, we successfully captured the subtleties and variations within the coastline region. This level of detail greatly enhances the realism and accuracy of the digital twin, providing users with a more immersive and educational experience. To develop a digital twin of the entire state of Florida, we first collected information from Google Earth. The extracted data underwent a processing step, resulting in the generation of 256x256 RGB-formatted square images that were tiled. These images were processed using various filters to address any color inconsistencies between the trained network and the test data, ensuring alignment with the training data. Figure 8 showcases the results of this technique, illustrating the model's capability to distinguish between different locations and accurately capture the aesthetic qualities of Florida's national parks. It is important to note that certain misclassifications in the images may arise due to discrepancies between the training data and the test data utilized in the model.

## 5. Discussion

We found that our approach to representing and classifying diverse terrain categories is effective based on the results of our study, including the generated global Terrain and Height map and the segmentation results obtained using the U-Net network. As well as demonstrating how deep learning methods can be used to develop digital twins of real-world locations, the successful modeling of the Florida coastal area using satellite imagery shows how significant it is. By combining advanced algorithms with high-quality data, we have achieved a comprehensive understanding of terrain characteristics and produced a detailed representation of the Florida coastal area. This digital twin serves as a valuable tool for a variety of applications, such as urban planning, environmental management, and disaster response. Our study also provides insights into the potential applications of the generated global Terrain and Height map. This map can be applied to a wide range of fields due to its artistic depiction of land types and accurate representation of elevation changes. In addition to providing cartographers and landscape artists with a realistic and visually appealing representation of global terrains, the map can also be used in virtual reality and gaming applications to create immersive and realistic virtual environments.

It has been demonstrated that deep learning methods can be effective in classifying terrain categories based on segmentation results obtained with the U-Net network. We were able to achieve high accuracy in segmenting different terrain types using neural networks and training the model on a diverse dataset. Land cover classification, environmental monitoring, and natural resource management are among the domains affected by this. For purposes such as land use planning, biodiversity assessment, and ecological modeling, accurate terrain segmentation can assist in identifying and monitoring specific land types. Additionally, the analysis of the segmentation performance using the ROC curve and the computation of AUC values have provided quantitative indications of the effectiveness of the model. It is evident from the high AUC values obtained that the model is capable of distinguishing between different types of terrain with high accuracy. This further strengthens the reliability of our approach and underscores its potential for real-world applications. The use of Google Earth satellite images has proven to be a successful strategy for modeling the Florida coastal area. In addition to providing valuable insight into the topography, vegetation, and infrastructure of the coastal region, the digital twin provides detailed representations of the area's coastline. In order to manage coastal resources, assess hazards, and plan for urban development, this information can be leveraged. The digital twin can be utilized by decision-makers to simulate and evaluate the potential impact of coastal erosion, sea level rise, and other environmental factors on the region. It is important to acknowledge that there are some limitations to our study. In the first instance, although we utilized a large dataset for training the U-Net network, the quality and coverage of the dataset can still influence the model's performance. It is likely that the generalization capabilities of the model will be enhanced if the dataset is enhanced by including more representative samples from different geographical regions. In addition, although a detailed representation of the Florida coastal area may be provided by the digital twin, it is important to note that it is merely a simplified and virtual representation of a real-life area, so there is an inherent level of uncertainty and approximation. As a result of the findings of this study, the field of terrain modeling and the development of digital twins can be further advanced in the future. In addition to improving the accuracy and efficiency of terrain segmentation, advanced deep-learning architectures and techniques may also be explored. The fidelity and realism of the digital twins can be enhanced by integrating additional data sources, such as LiDAR data or multispectral satellite imagery. Additionally, incorporating dynamic elements, such as real-time environmental data and simulation of natural processes, can enhance the complexity and utility of the digital twins.

## 6. Conclusion

The application of digital twins, as presented in our study of the Florida coastal area, holds immense potential across various domains. Digital twins provide valuable insights and tools for decision-making and analysis in a wide range of areas, including environmental planning, environmental management, disaster preparation, and tourism. Through the use of digital twins, stakeholders can gain a more comprehensive understanding of the region, make informed decisions, and contribute to sustainable development. The digital twin of the Florida coastal area serves as a testament to the wide-ranging applications and benefits of digital twin technology in real-world contexts. In conclusion, our study has demonstrated the effectiveness of deep learning methods in generating a global Terrain and Height map, as well as in accurately classifying different terrain categories utilizing the U-Net network. Creating a detailed digital twin of the Florida coastline further illustrates the potential of these techniques to develop virtual representations of real-world locations. Through this research, insights can be gained to improve land cover classification, environmental monitoring, and urban planning efforts, ultimately contributing to more informed decision-making and sustainable development.